\documentclass[fleqn,12pt,twoside]{article}
\usepackage{espcrc1}
\usepackage{epsfig}

\usepackage{graphicx}
\usepackage[figuresright]{rotating}

\newcommand{\AmS}{{\protect\the\textfont2
  A\kern-.1667em\lower.5ex\hbox{M}\kern-.125emS}}

\def\numu{\nu_\mu}

\title{CNGS beam monitor with the LVD detector.}

\author{
M.Aglietta\address[a]{Institute of Physics of Interplanetary Space, IFSI, CNR, Torino, University of Torino \\and INFN-Torino, Italy,},
P.Antonioli\address[b]{University of Bologna and INFN-Bologna, Italy,},
G.Bari\addressmark[b],
C.Castagnoli\addressmark[a],
W.Fulgione\addressmark[a],
P.Galeotti\addressmark[a],
M.Garbini\addressmark[b],
P.L.Ghia\addressmark[a],
P.Giusti\addressmark[b],
E.Kemp\address[c]{University of Campinas, Campinas, Brazil,},
A.S.Malguin\address[d]{Institute for Nuclear Research, Russian Academy of Sciences, Moscow, Russia,},
H.Menghetti\addressmark[b],
A.Pesci\addressmark[b],
I.A.Pless\address[e]{Massachusetts Institute of Technology, Cambridge, USA,},
A.Porta\addressmark[a],
V.G.Ryasny\addressmark[d],
O.G.Ryazhskaya\addressmark[d],
O.Saavedra\addressmark[a],
G.Sartorelli\addressmark[b]\thanks{Corresponding author: Gabriella Sartorelli, c/o Dipartimento di Fisica, via Irnerio 46, 40126 Bologna - Italy, sartorelli@bo.infn.it}, 
M.Selvi\addressmark[b]\thanks{Corresponding author: Marco Selvi, c/o Dipartimento di Fisica, via Irnerio 46, 40126 Bologna - Italy, selvi@bo.infn.it},
C.Vigorito\addressmark[a],
L.Votano\address[f]{INFN-LNF, Frascati, Italy},
V.F.Yakushev\addressmark[d],
G.T.Zatsepin\addressmark[d],
A.Zichichi\addressmark[b]
}

\begin{document}

\maketitle

\begin{abstract}
Abstract:\\
The importance of an adequate CNGS beam monitor at the Gran Sasso Laboratory has been stressed in many papers.
Since the number of internal $\nu_\mu$ CC and NC interactions in the various detectors will not allow to collect statistics rapidly,
one should also be able to detect the 
$\nu_\mu$ CC interactions in the upstream rock.\\
In this study we have investigated the performances of the LVD detector as a monitor for the CNGS neutrino beam.\\
Thanks to its wide area ($13 \times 11~m^2$ orthogonal to the beam direction) LVD can detect about $120$ muons per day originated by $\nu_\mu$ CC interactions in the rock.\\ 
The LVD total mass is $\sim2~kt$. This allows to get 30 more CNGS events per day as internal $(NC~+~CC)$ $\nu_\mu$ interactions, for a total of $\sim 150$ events per day. A $3\%$ statistical error can be reached in 7 days. Taking into account the time characteristics of the CNGS beam, the cosmic muon background can be reduced to a negligible level, of the order of $1.5$ events per day.\\
\end{abstract}


\newpage

\section{Introduction}
The CNGS beam from CERN to the Gran Sasso Laboratory (LNGS) is a wide-band high-energy $\nu_\mu$ beam  ($<E>\sim 23~GeV$) optimized for $\tau$ appearance. It provides $\sim 2600 ~CC / kt / y$ and $\sim 800 ~NC / kt / y$ at Gran Sasso \cite{cngs2000}, that is, assuming $200$ days of beam-time per year, a total number of $\sim17~ (CC+NC)/kt/day$.

In order to provide an adequate monitoring of the beam performance it has been estimated \cite{cngs2001} that one should be able to collect an event sample affected by a statistical error of the order of $3\%$ in a few days time. In table \ref{ta:time} the number of days needed in order to get this statistical significance is shown as a function of the total number of CNGS events collected per day at Gran Sasso.

\vspace{.2 cm} 
\begin{table}[h]
\begin{center}
\begin{tabular} {|c|c|} 
\hline
\hline
CNGS events detected per day & Nb. of days needed to get $3\%$ statistical error\\
\hline
$50$  & $22$ \\
$100$  & $11$ \\
$200$  & $5.5$ \\
$500$  & $2.2$ \\
$1000$ & $1.1$ \\
\hline \hline
\end{tabular} 
\vspace{.5 cm} 
\caption{Number of days needed to get $3\%$ statistical error as a function of the number of CNGS events detected per day.}
\label{ta:time}
\end{center}
\end{table}

Even considering an overall mass of the various experiments which will be active in GS at the beam start up as large as $5~kt$, the number of CC and NC interaction per day, internal to the detectors, will be only $\sim~80$ and therefore more than $14$ days will be needed to collect a sample with the statistical significance mentioned above.

The number of events observed per day can be increased considering the muons produced by the $\nu_\mu$ CC interactions in the upstream rock, emerging into the experimental halls and detected either by a simple wide area dedicated monitor or by the running experiments.

The LVD experiment is primarily dedicated to the search for neutrinos from gravitational stellar collapses and has been running since 1991.  
In this study, we have investigated the beam monitor capabilities of the existing LVD detector \cite{LVDdescr}, whose beam orthogonal surface is $13 \times 11~m^2$, larger than the other foreseen CNGS experiments.
A detailed MonteCarlo simulation to estimate both the muon flux inside LVD and the number of detectable CC and NC internal interactions has been developed.

\section{Event Generation}
\label{pa:evgen}
\begin{figure}[htbp]
\begin{center}
\epsfig{file=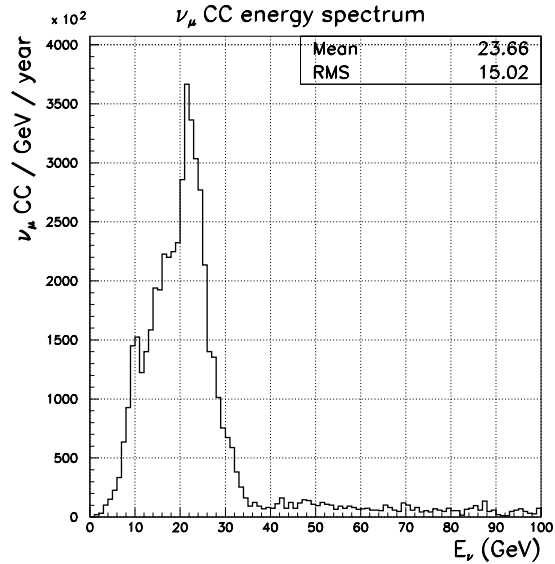,width=8cm,height=8cm}
\caption {Charged current $\numu$ interactions energy spectrum of CNGS.} 
\label{fi:numuspec}
\end{center}
\end{figure}

\begin{figure}[htbp]
\begin{center}
\epsfig{file=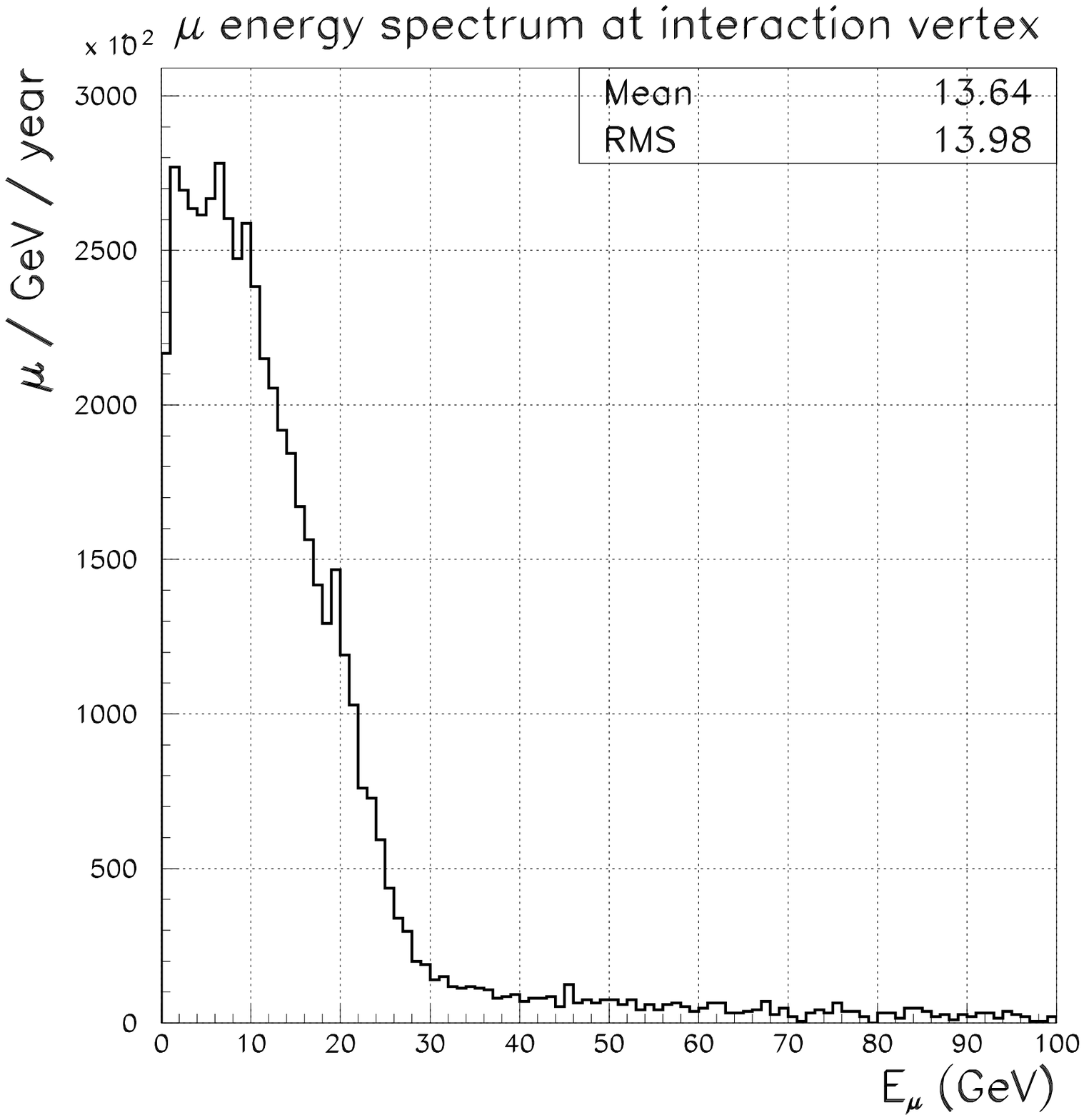,width=.45\linewidth}
\epsfig{file=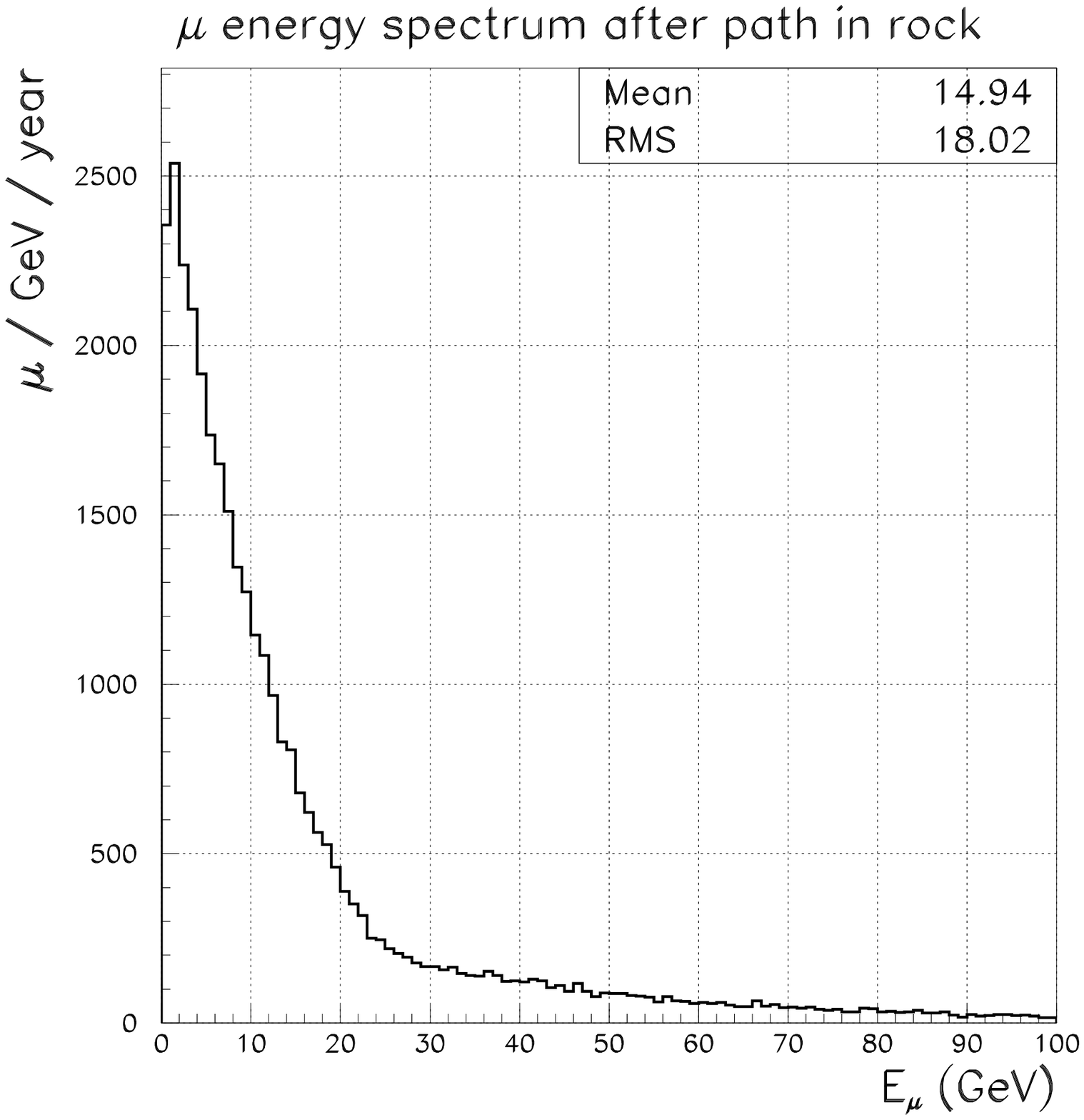,width=.45\linewidth}
\caption {Energy spectra of the $\mu$ generated in  charged current $\numu$ interactions at the interaction vertex ({\it left}) and after path in rock, when entering the experimental hall ({\it right}).} 
\label{fi:muspec}
\end{center}
\end{figure}

The charged current interactions of the CNGS beam neutrinos have been
simulated using the Lipari generator \cite{lipgen} in which neutrino interactions are calculated 
with GRV94 parton distributions \cite{GRV94} with explicit inclusion 
of the contribution of quasi-elastic scattering and of single pion 
production to the neutrino cross-sections.
The neutrino energy has been sampled from the reference CNGS beam  
spe\-ctrum~\cite{cngs2000}. 
In fig.~\ref{fi:numuspec} is shown the energy spectrum of $\nu_\mu$ CC interactions, cross section weighed.
The muon energy at the interaction vertex is shown in figure \ref{fi:muspec} ({\it left}).

\subsection{Muons generated in the rock}

\begin{figure}[htbp]
\begin{center}
\epsfig{file=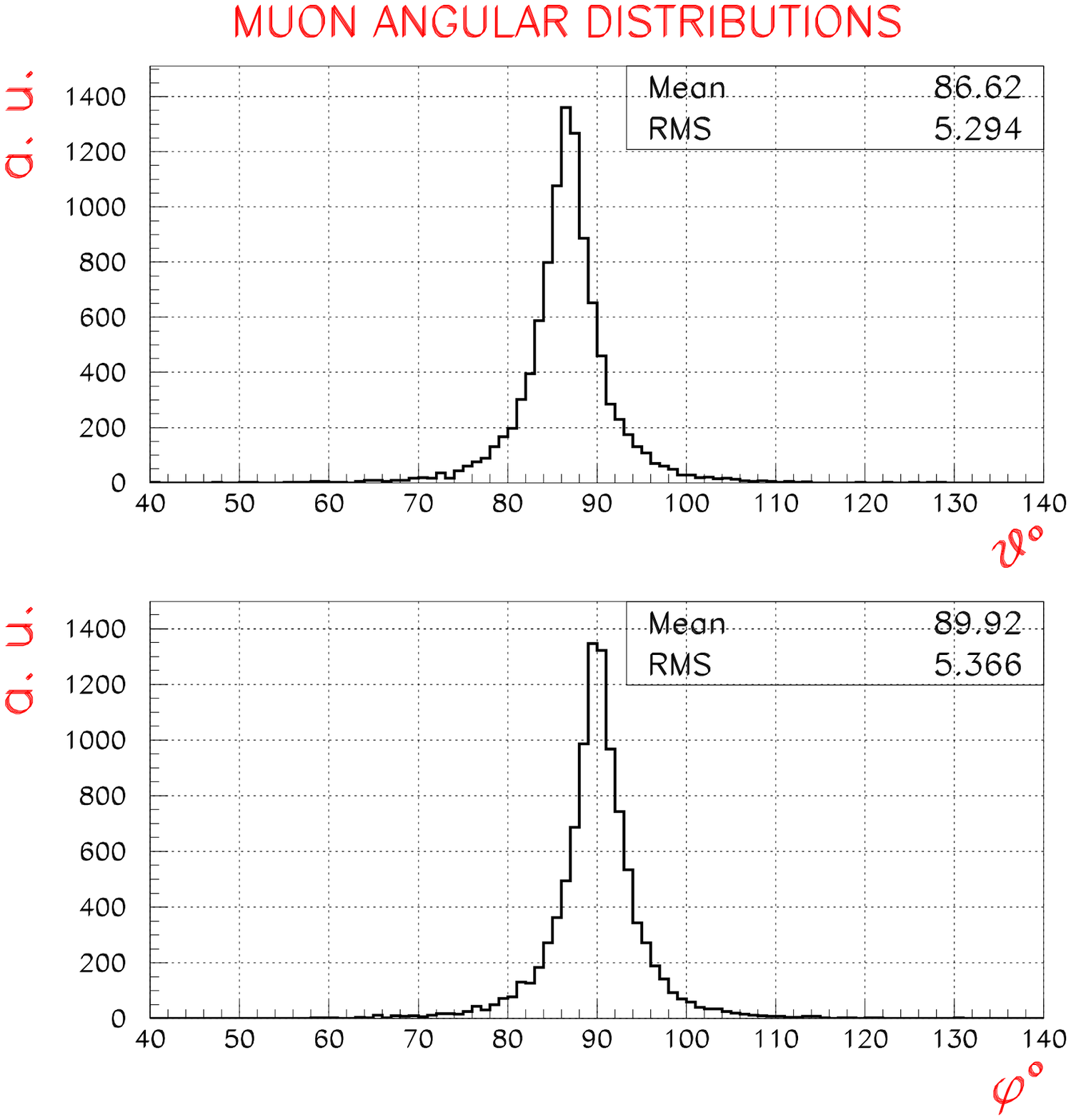,width=.7\linewidth}
\caption {Zenithal ({\it top}) and azimuthal ({\it bottom}) angle distributions of {\it CNGS muons} in the LVD reference system: $\theta=90^o \equiv$ horizontal; $\phi=90^o \equiv$ parallel to the hall A axis.} 
\label{fi:thephi}
\end{center}
\end{figure}

The muons originated in the rock by the beam neutrinos ({\it CNGS muons}), which have an angle of  $3.2^\circ$ (upward) from the horizon, have been propagated through the rock with MUSIC \cite{music}, a three dimensional code which takes into account multiple scattering, energy losses due to ionization, bremsstrahlung, pair production and nuclear interactions. The muon energy spectrum after emerging from the rock is shown in figure  \ref{fi:muspec} ({\it right}), while the zenithal and azimuthal angle in the LVD reference system are shown in figure \ref{fi:thephi}.

The CNGS neutrino interactions, considered for the present study, are generated in a rock cylinder of height $370~m$ and radius $25~m$, just in front of the LNGS Hall A. The cylinder and Hall A axis coincide. The generation volume and mass are thus $726493~m^3$ and $1969~ kt$.
The muon range in the rock, in a first approximation, is given by $$ R = \frac{E_\mu}{  \rho_{GS} \times \frac{dE}{dX}} $$ where the Gran Sasso rock density $\rho_{GS}=2.71~g/cm^3$ and $\frac{dE}{dX}=2~MeV/(g/cm^2)$. 
We assumed a maximum muon energy of $200~GeV$ that corresponds to a $\sim 370~m$ path in GS rock. This explains the choice of the cylinder height. We use a large radius ($25~m$ with respect to the $13 \times 11 ~ m^2$ of LVD front surface) in order to take into account also muons originated 
far from the detector axis, but able to reach it because of their large angle (with respect to the beam direction).\\

The interaction vertex is selected uniformly in the cylinder volume.
We define ``Entrance LNGS ({\bf EL}) Plane'' the infinite vertical plane separating the rock cylinder face  and the entrance wall of the experimental hall. The muon propagation from the interaction vertex to the LVD detector is subdivided into two main steps:
\begin{itemize}
\item The muons, generated with energy distribution according to figure \ref{fi:muspec} ({\it left}), are traced by MUSIC until they lose all their energy or until they reach the EL plane. The muons emerging from the EL plane with non-zero energy are $6.8\%$ of the generated events.
\item The event is taken into account if the emerging muon direction points to a mother volume ($13.75~m~ (horiz.)  \times  12~m~ (vert.) \times~ 24.8~m~ (lenght)$) which encloses the LVD detector and if the muon has enough energy  to reach the LVD mother volume from the emerging point in the EL plane. This request is satisfied by $9.6\%$ of the muons that emerge from the EL plane.
\end{itemize}

For comparison purposes we have calculated the total number of $\nu_\mu$ CC interactions in one CNGS year (200 days) inside the rock volume using the number of $CC/proton~ on~ target~ (pot)/kt$ and $pot/year$ given in ref. \cite{cngs2000}.

Accordingly,
the number of muons hitting the LVD mother volume
in one year is thus $\sim 33600$, i.e. $\sim 170$ per day.

A summary of the elements needed for the muon number calculation in LVD is shown in table \ref{cuts}.

\vspace{.2 cm}
\begin{table}[h]
\begin{center}
\begin{tabular} {|l|c|} 
\hline
\hline
Rock cylinder Volume & $726493~m^3$ \\
\hline
Rock cylinder Mass & $1969~ kt$ \\
\hline
Nominal CNGS beam intensity & $4.5 \times 10^{19} pot/year$ \\
\hline
CC Interaction Probability & $5.85 \times 10^{-17}~CC/pot/kt$ \\
\hline
$\nu_\mu$ CC interactions in the rock & $5.18 \times 10^6~ per~ year$ \\
\hline
$\mu$ survival probability at the EL plane  & $6.8\%$ \\
\hline
Nb. of $\mu$ at the EL plane & $351600~ per~ year$ \\
\hline
Probability to hit LVD ``mother volume'' & $9.6\%$ \\
\hline 
Nb. of $\mu$ hitting LVD ``mother volume'' & $33600~ per~ year$\\
\hline \hline
\end{tabular} 
\vspace{.5 cm}
\caption{Number of muon detectable in LVD.}
\label{cuts}
\end{center}
\end{table}

\subsection{Internal CC and NC events} 
The LVD detector consists of 912 counters constituted of liquid scintillator in stainless steel tanks. 8 counters are grouped in a supporting structure (portatank) and 38 portatanks are arranged in a compact structure to form a LVD tower.

The total active scintillator mass is $\sim 1050~t$ while the total mass of stainless steel tank and portatank, that can efficiently act as high energy neutrino target, is $\sim 770~t$. The interaction vertex of CC and NC internal events has been distributed uniformly in the apparatus. The number of events generated in steel and in scintillator is proportional to the relative weight of the two components.

The number of CC and NC events, at nominal beam intensity,
are respectively $4770$ and $1460$ per year, as shown in table \ref{cutsn}. The number of CC interactions is calculated accordingly to ref. \cite{cngs2000} while the ratio between the number of CC and NC interactions is calculated accordingly to the event generator described in paragraph \ref{pa:evgen}.

\vspace{.2 cm}
\begin{table}[h]
\begin{center}
\begin{tabular} {|l|c|c|} 
\hline
\hline
 & {\bf Volume} & {\bf Mass} \\
\hline
Scintillator & $1340~m^3$ & $1044~ t$ \\
\hline
Stainless Steel & $98.5~m^3$ & $770~ t$ \\
\hline
Total Target  & & $\sim 1810 ~t$ \\
\hline
\hline
 & {\bf CC} & {\bf NC} \\
\hline
$\nu_\mu$ interactions in LVD  & $4770~ per~ year$ & $1460~ per~ year$ \\
\hline
Total nb. of internal events  &\multicolumn{2}{c}{$6230~per~year$} \\
\hline \hline
\end{tabular} 
\vspace{.5 cm}
\caption{Number of internal CC and NC events in LVD.}
\label{cutsn}
\end{center}
\end{table}

\section{LVD response}
The detailed detector geometry and the particle interactions with the detector material have been taken into account with a GEANT simulation of the LVD detector (see \cite{tesiMS} for a detailed description of the simulation code). 

\subsection{Event Selection: muons generated in the rock}

\begin{figure}[htbp]
\begin{center}
\epsfig{file=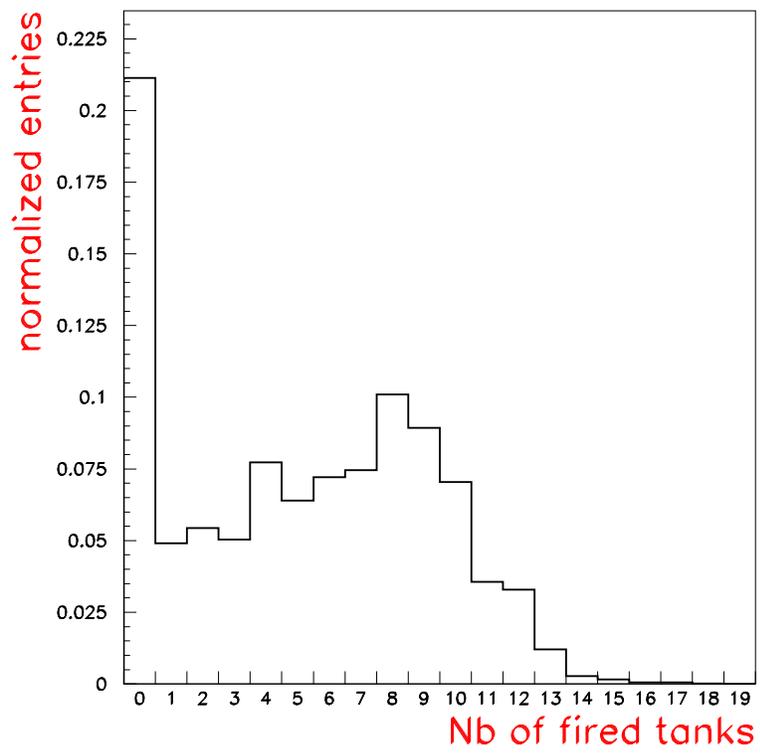,width=.7\linewidth}
\caption{Percentage of counters hit by {\it CNGS muons} per event.}
\label{fi:fired}
\end{center}
\end{figure}

\begin{figure}[htbp]
\begin{center}
\epsfig{file=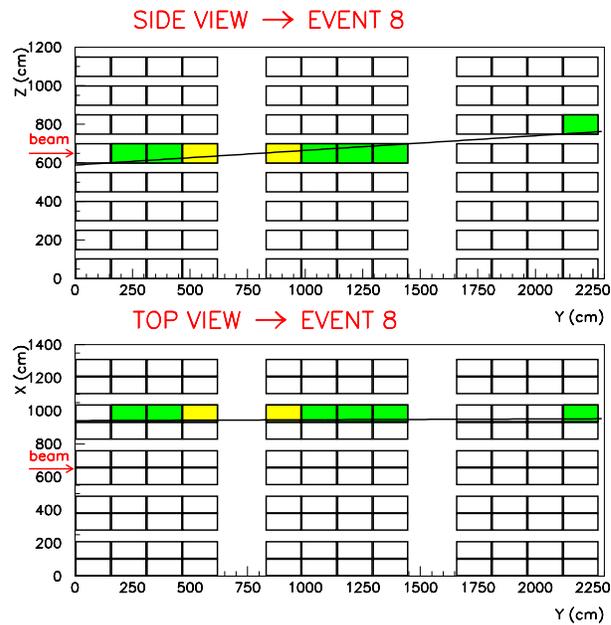,width=.55\linewidth}
\caption{Event display of a {\it CNGS $\mu$} crossing 8 scintillator counters.}
\label{fi:display}
\end{center}
\end{figure}

\begin{figure}[htbp]
\begin{center}
\epsfig{file=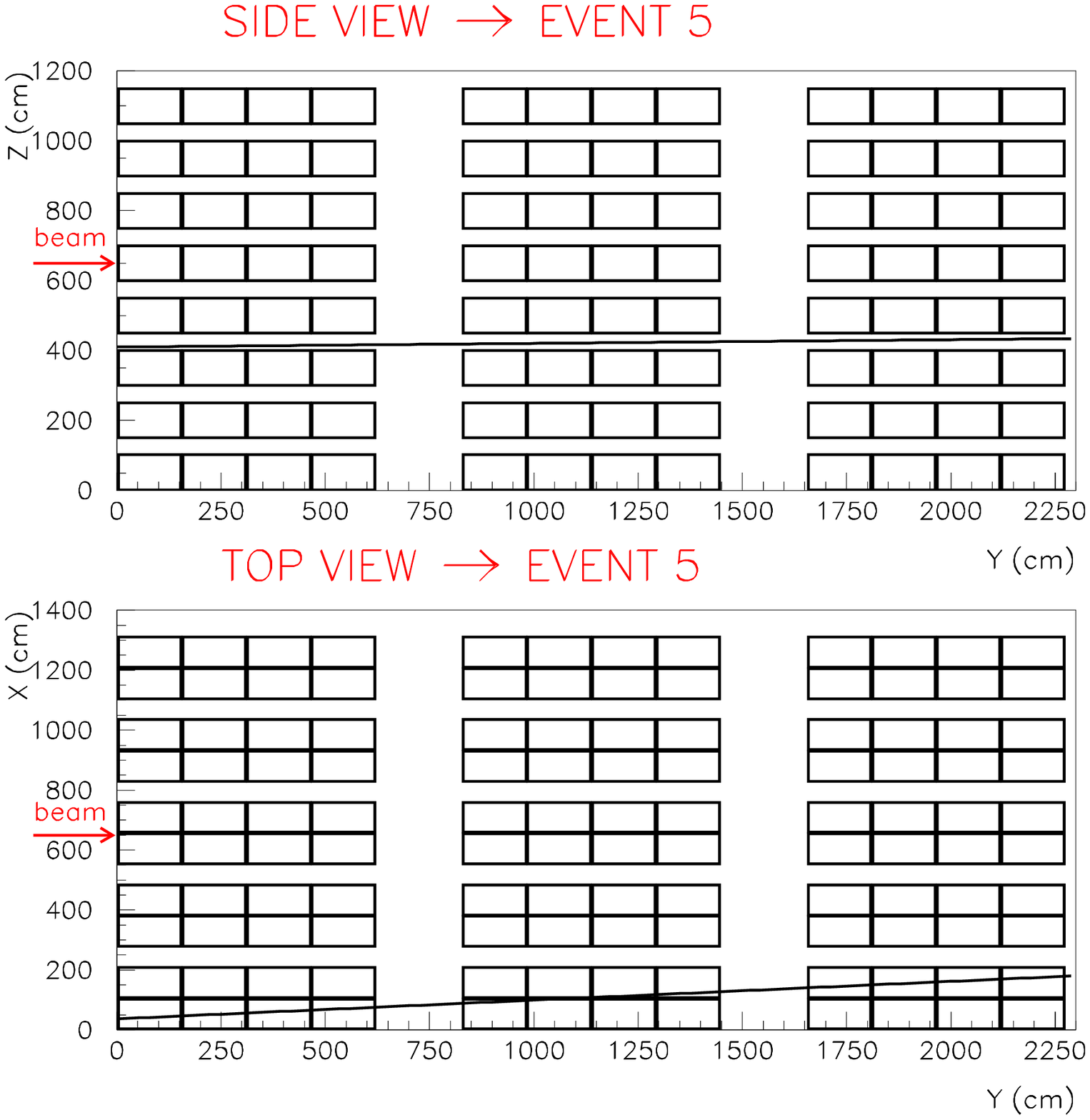,width=.55\linewidth}
\caption{Event display of a {\it CNGS $\mu$} passing through a corridor.}
\label{fi:displayb}
\end{center}
\end{figure}

According to the simulation described in paragraph \ref{pa:evgen}, $87\%$ of the generated muon tracks hit the mother volume in the front face, while the remaining $13\%$ enter LVD in the lateral faces (left, right, top, bottom).\\
Figure \ref{fi:fired} shows the distribution of the number of counters hit by the muons emerging from the rock.
About $21\%$ of them pass through the detector dead regions (corridors), and the corresponding geometrical efficiency is about $79\%$.
See some event displays in figure \ref{fi:display} and \ref{fi:displayb}.

\begin{figure}[htbp]
\begin{center}
\epsfig{file=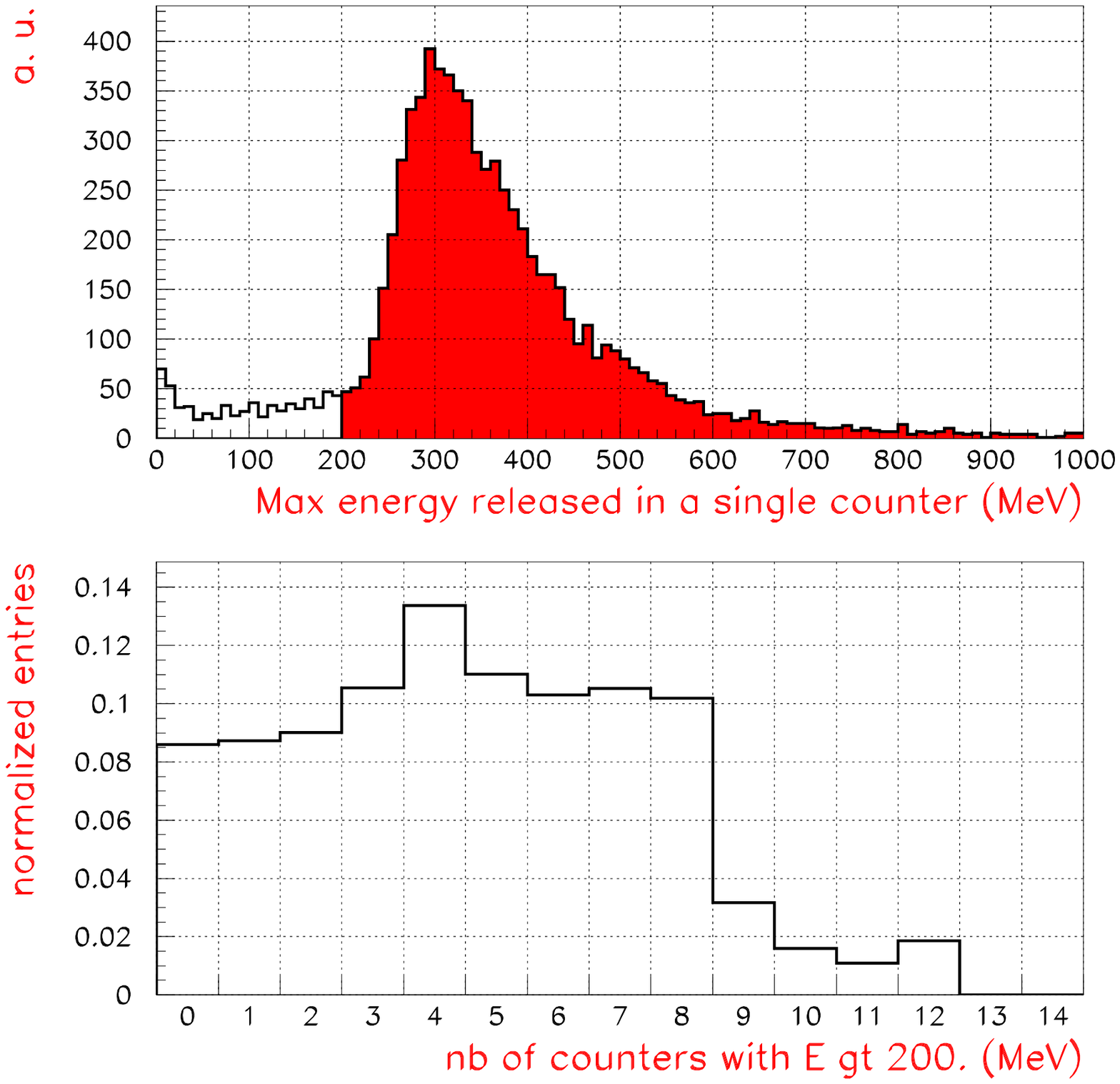,width=.7\linewidth}
\caption {{\it Top: }Maximum energy loss in a single scintillation counter by a {\it CNGS $\mu$}. {\it Bottom: } Number of scintillation counters with an energy release greater than $200~MeV$.} 
\label{fi:defmu}
\end{center}
\end{figure}

Each scintillation counter is $1~m\times 1~m$ of front area and $1.5~m$ along the beam direction.
In figure \ref{fi:defmu} ({\it top}) the maximum energy released by muons in one single counter per event is shown. Figure \ref{fi:defmu} ({\it bottom}) shows the number of counters with an energy release greater than $200~MeV$, the entries being normalized to the total number of events with at least one fired counter.
Assuming, as an event selection criteria, the presence of at least one counter with an energy loss greater than $200~MeV$, we are able to detect $92\%$ of the total number of muons entering an active part of the LVD detector (detection efficiency). The global efficiency (geometrical + detection) is thus $72\%$.

Therefore the mean number of detected muons in LVD, at the nominal CNGS beam intensities, is about $24200$ per year, i.e. $120$ per day. 

\vspace{.2 cm}
\begin{table}[h]
\begin{center}
\begin{tabular} {|l|c|} 
\hline
\hline
Nb. of $\mu$ hitting LVD ``mother volume'' & $33600~ per~ year$\\
\hline
Geometrical efficiency & $79\%$ \\
\hline
Nb. of $\mu$ hitting LVD sensitive volume & $26200~ per~ year$\\
\hline
Selection cut efficiency & $92\%$ \\
\hline
Nb. of detected $\mu$ & $24200 ~ per~ year$ \\
\hline \hline
\end{tabular} 
\vspace{.5 cm}
\caption{Number of {\it CNGS muons} detected in LVD.}
\label{cuts2}
\end{center}
\end{table}

\subsection{Event selection: internal CC and NC events} 
\begin{figure}[htbp]
\begin{center}
\epsfig{file=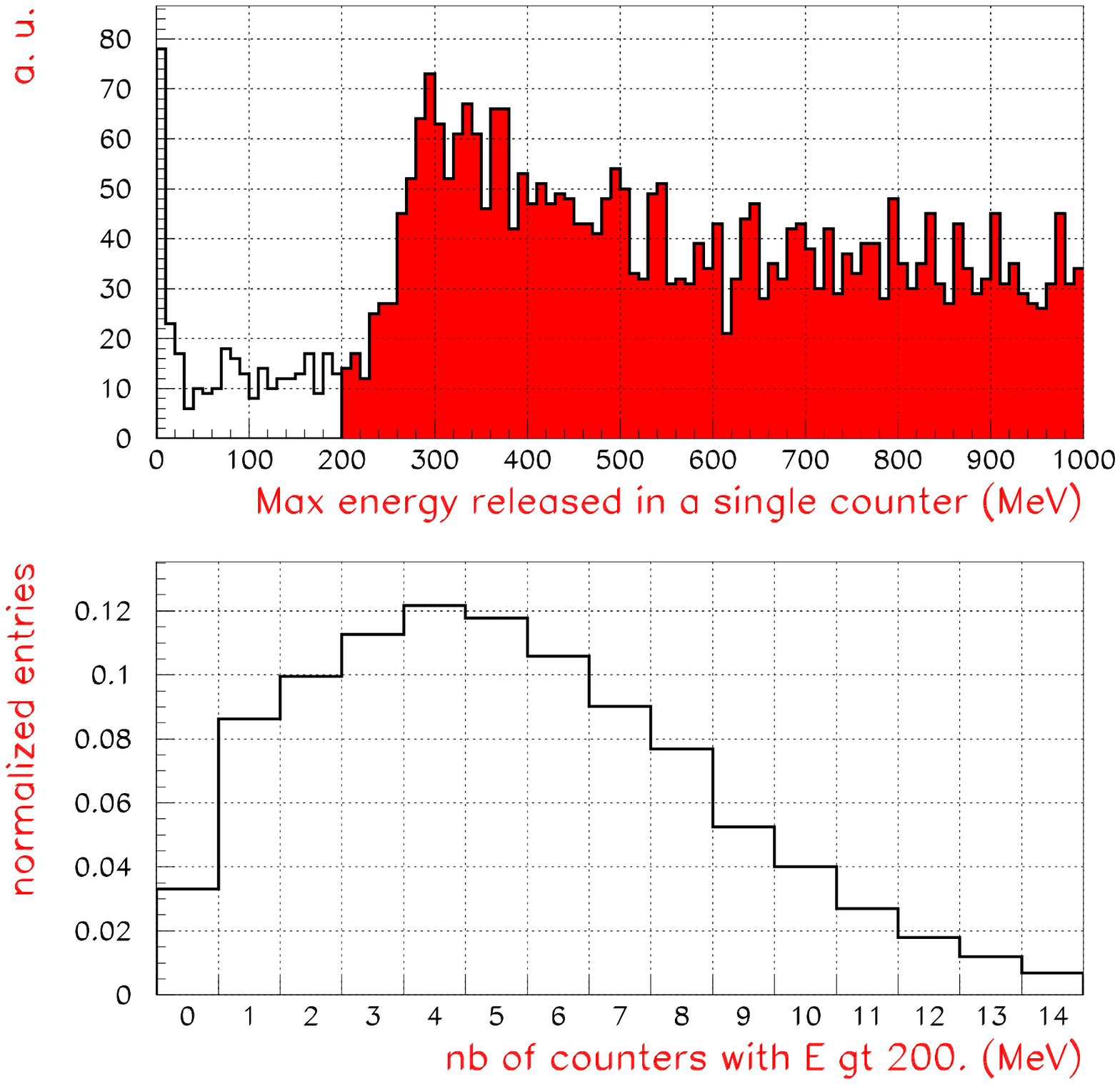,width=.45\linewidth}
\caption {{\it Top: }Maximum energy loss in a single scintillation counter by internal $\nu_\mu$ CC interactions. {\it Bottom: } Number of counters with an energy release greater than $200~MeV$.} 
\label{fi:defmucc}
\end{center}
\end{figure}

\begin{figure}[htbp]
\begin{center}
\epsfig{file=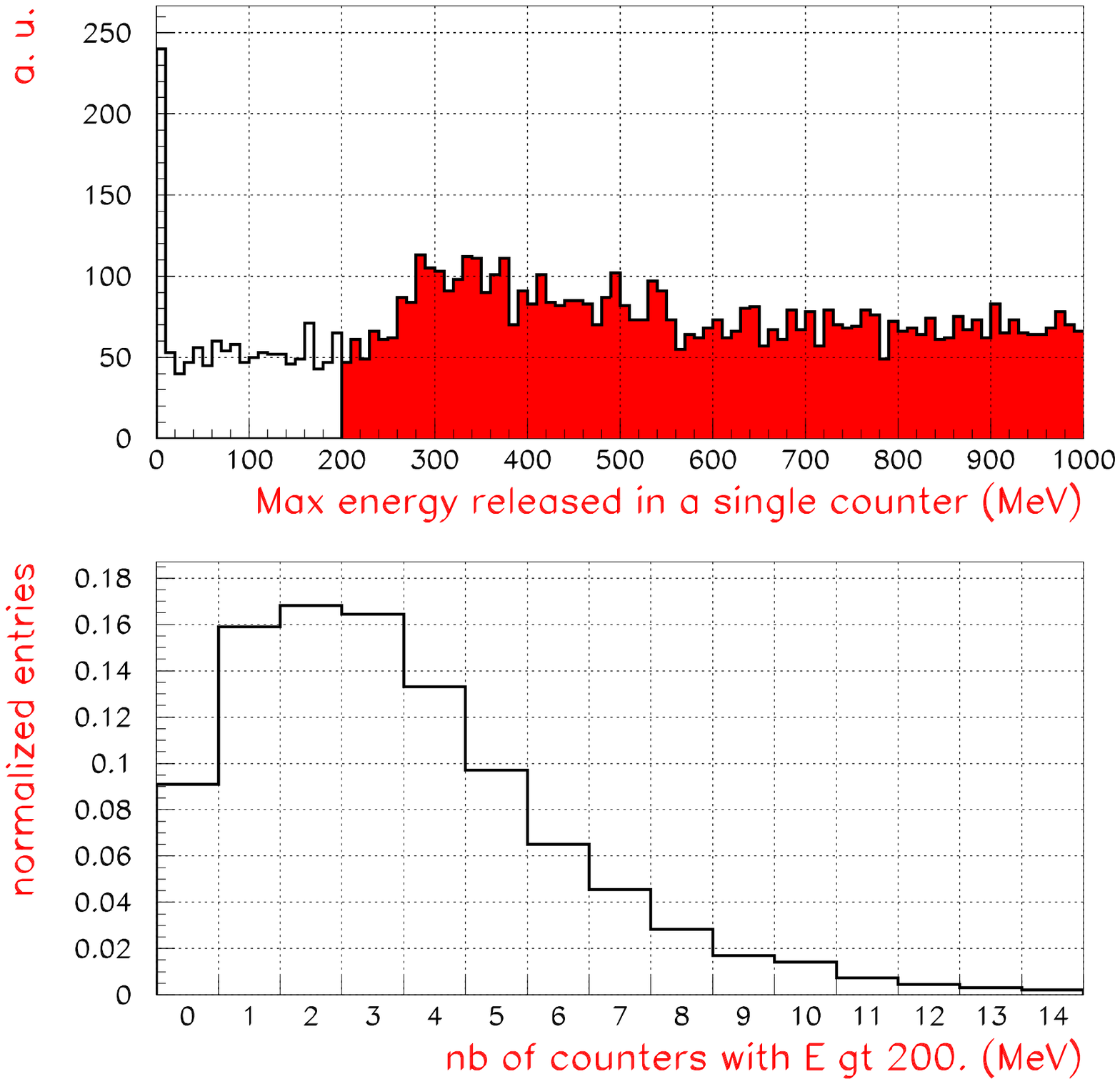,width=.45\linewidth}
\caption {{\it Top: }Maximum energy loss in a single scintillation counter by internal $\nu_\mu$ NC interactions. {\it Bottom: } Number of counters with an energy release greater than $200~MeV$.} 
\label{fi:defmunc}
\end{center}
\end{figure}

\begin{figure}[htbp]
\begin{center}
\epsfig{file=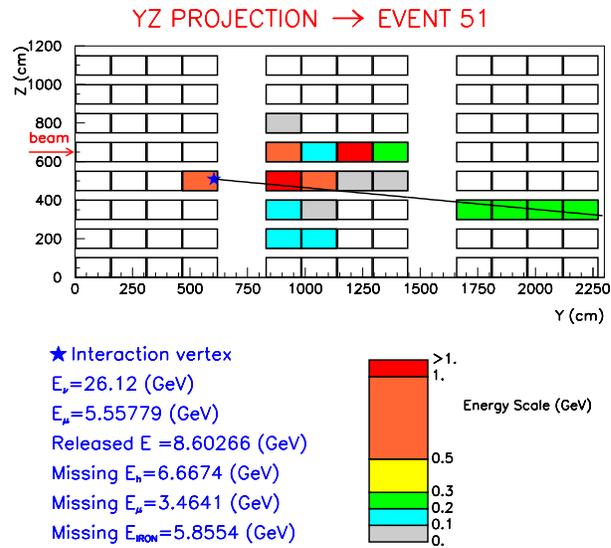,width=.55\linewidth}
\caption{Event display of CNGS $\nu_\mu$ charged current internal interaction.}
\label{fi:dispcc}
\end{center}
\end{figure}

\begin{figure}[htbp]
\begin{center}
\epsfig{file=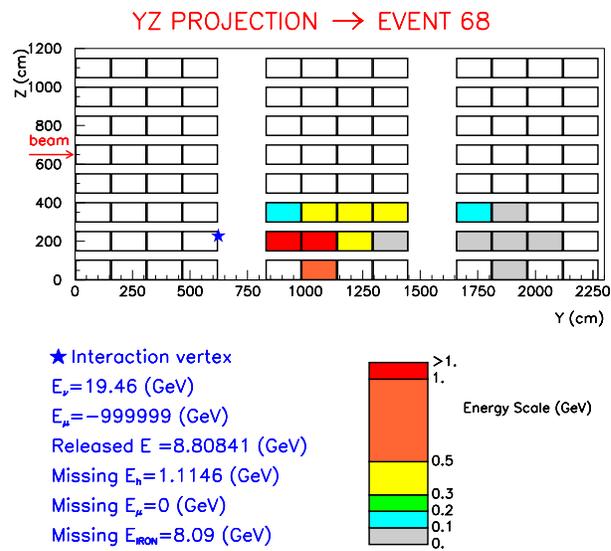,width=.55\linewidth}
\caption{Event display of CNGS $\nu_\mu$ neutral current internal interaction.}
\label{fi:dispnc}
\end{center}
\end{figure}

In the case of neutrinos interacting inside the LVD detector, all the particles generated (muons and hadrons) are traced through the apparatus and the energy released in the scintillation counters is the sum of the contribution from all the traversing particles. Applying the same selection cut described in the previous paragraph (at least one counter with $E > 200~ MeV$), 
the efficiency for the CC (NC) internal event sample is $97\%$ ($91\%$), as shown in figures \ref{fi:defmucc} and \ref{fi:defmunc}.  
This gives $\sim 4630$ CC and $\sim 1330$ NC events in one year, corresponding to $\sim 30$ additional CNGS events per day.
Two examples of internal $\nu_\mu$ CC and NC interaction event displays are shown in figure \ref{fi:dispcc} and \ref{fi:dispnc}.

\vspace{.2 cm}
\begin{table}[h]
\begin{center}
\begin{tabular} {|l|c|c|} 
\hline
\hline
 & CC & NC \\
\hline
$\nu_\mu$ interactions in LVD  & $4770~ per~ year$ & $1460~ per~ year$ \\
\hline
efficiency for the selection cut & $97\%$ & $91\%$\\
\hline
Nb. of detected events & $4630~ per~ year$ & $1330~ per~ year$ \\
\hline
Total nb. of detected events  &\multicolumn{2}{c}{$5960~per~year$} \\
\hline \hline
\end{tabular} 
\vspace{.5 cm}
\caption{Number of internal CC and NC events detected in LVD.}
\label{cutsn2}
\end{center}
\end{table}

\section{Total event rate and background}
We stress the point that here we are not interested in distinguishing between neutrino interacting in the rock events
and internal neutrino events; the loose cut chosen has the advantage of a very good efficiency needed for a good monitoring task.  
Taking into account both the muons from CC interactions in the rock and the internal CC and NC interactions, the statistical error obtained during various number of days of run is shown in table \ref{ta:staterr}.

\vspace{.2 cm}
\begin{table}[h]
\begin{center}
\begin{tabular} {|l|c|} 
\hline
\hline
Run duration & Statistical error \\
\hline
1 day & $8.2\%$ \\
4 days & $4.1\%$ \\
7 days & $3.1\%$ \\
10 days & $2.6\%$ \\
\hline \hline
\end{tabular} 
\vspace{.5 cm}
\caption{Statistical error on the number of CNGS events in LVD.}
\label{ta:staterr}
\end{center}
\end{table}

\newpage

The main background source is due to cosmic muons. 
The rate in the whole LVD detector is about 9300 muons/day (about $6.5$ per minute), considering the events with at least one scintillation counter fired by the muon. 
The requirement of at least one counter with an energy loss greater than 200 MeV rejects $20\%$ of them, 
leaving about 7500 muons per day. Considering the $10^4$ reduction factor due to the CNGS beam timing characteristics ($10.5~ \mu s$ of spill length and $50~ ms$ inter-spill gap \cite{beamspill}), the actual number of background $\mu$ per day is $\sim 1.5$, practically negligible.

\section{Comparison with a dedicated muon track detector}
In order to evaluate the capabilities of a dedicated muon monitor at LNGS, we used the same software chain  for the event generation and the muon propagation in rock. 
Thereafter we simulate the event in the ideal detector.\\
We define the muon monitor as three vertical planes made of RPC's (or Limited Streamer Tubes) with $100\%$ efficiency, separated by $1~m$ from each other along the beam direction. A muon is tagged when three planes are crossed by the muon track. At this level we don't add any consideration about the cosmic muon background rejection of this kind of detector, that must be studied in details. The number of muons detected per day by the detectors are shown in table \ref{ta:rpc} for various trasversal dimensions. We remark that the total number of CNGS interaction detected by LVD (muons from the rock + internal events) is similar to the number of muon seen by the biggest dedicated muon monitor.

\vspace{.2 cm}
\begin{table}[h]
\begin{center}
\begin{tabular} {|l|c|} 
\hline
\hline
Detector & Number of detected CNGS events \\
\hline
$13m \times 13m$ (as proposed in \cite{cngs2001}) & $146$ muons/day \\
$13m \times 14.5m$ (in the biggest hall C) & $165$ muons/day \\
$8m \times 9m$   & $63$ muons/day  \\
\hline
LVD &  $120$ muons/day (+ about $30$ internal events)  \\
\hline \hline
\end{tabular} 
\vspace{.5 cm}
\caption{Number of muons detected per day by a dedicated muon track detector.}
\label{ta:rpc}
\end{center}
\end{table}

\section{Summary and conclusions}
The importance of a CNGS beam monitor apparatus in the experimental halls of Gran Sasso Laboratory has been stressed in many papers \cite{cngs2000} \cite{cngs2001}.
In this paper the performances of the LVD detector as a monitor for the CNGS neutrino beam are shown. \\
Thanks to its wide area ($13 \times 11~m^2$ orthogonal to the beam direction) LVD can detect about $120$ muons per day using a very simple selection cut (at least one counter with an energy loss greater than $200~MeV$). With a $\sim2~kt$ total mass, LVD could detect 30 more CNGS events per day as internal $(NC~+~CC)$ $\nu_\mu$ interactions, for a total of $\sim 150$ events per day. A $3\%$ statistical error can thus be reached in 7 days (see table \ref{ta:staterr}).   The cosmic muons background can be reduced to a negligible level, of the order of about one event per day, by taking into account the CNGS beam spill.\\
These results compare well with the expected performances of a dedicated muon detector.

\section{Acknowledgements}                   
This work is supported by the Italian Institute for Nuclear Physics (INFN) and in part by the Italian Ministry of Education, University and Scientific Research (MIUR) and the Russian Foundation of Basic Research (RFBR Grant 03-02-16413). 

\end{document}